\documentclass[iop,apj]{emulateapj}

\usepackage{acronym}
\usepackage{booktabs}
\usepackage{dcolumn}
\usepackage{graphicx}
\usepackage{amsmath,amsfonts,amssymb}
\usepackage{mathrsfs}
\usepackage[utf8]{inputenc}
\usepackage[normalem]{ulem}

\usepackage[breaklinks,colorlinks,citecolor=blue]{hyperref}

\newcommand{\myhead}[1]{\multicolumn{1}{c}{#1}}
\newcolumntype{d}[1]{D{.}{.}{#1}}

\usepackage{natbib}
\newcommand{\GW}{GW170817}
\newcommand{\AT}{AT2017gfo} 

\usepackage{color}

\newacro{ADM}{Arnowitt-Deser-Misner}
\newacro{AMR}{adaptive mesh-refinement}
\newacro{BH}{black hole}
\newacro{BBH}{binary black-hole}
\newacro{BHNS}{black-hole neutron-star}
\newacro{BNS}{binary neutron star}
\newacro{CCSN}{core-collapse supernova}
\newacroplural{CCSN}[CCSNe]{core-collapse supernovae}
\newacro{CMA}{consistent multi-fluid advection}
\newacro{DG}{discontinuous Galerkin}
\newacro{HMNS}{hypermassive neutron star}
\newacro{EM}{electromagnetic}
\newacro{ET}{Einstein Telescope}
\newacro{EOB}{effective-one-body}
\newacro{EOS}{equation of state}
\newacroplural{EOS}[EOSs]{equations of state}
\newacro{FF}{fitting factor}
\newacro{GR}{general relativity}
\newacro{GRLES}{general-relativistic large-eddy simulation}
\newacro{GRHD}{general-relativistic hydrodynamics}
\newacro{GRMHD}{general-relativistic magnetohydrodynamics}
\newacro{GW}{gravitational wave}
\newacro{ILES}{implicit large-eddy simulations}
\newacro{LIA}{linear interaction analysis}
\newacro{LES}{large-eddy simulation}
\newacroplural{LES}[LES]{large-eddy simulations}
\newacro{MRI}{magnetorotational instability}
\newacro{NR}{numerical relativity}
\newacro{NS}{neutron star}
\newacro{PNS}{protoneutron star}
\newacro{SASI}{standing accretion shock instability}
\newacro{SGRB}{short $\gamma$-ray burst}
\newacro{SN}{supernova}
\newacroplural{SN}[SNe]{supernovae}
\newacro{SNR}{signal-to-noise ratio}

\begin{document}

\title{\GW{}: Joint Constraint on the Neutron Star Equation of State\\
from Multimessenger Observations}

\author{David Radice\altaffilmark{1,2}, Albino
Perego\altaffilmark{3,4,5}, Francesco Zappa\altaffilmark{5},
and Sebastiano Bernuzzi\altaffilmark{5,3}.}
\altaffiltext{1}{Institute for Advanced Study, 1 Einstein Drive,
Princeton, NJ 08540, USA}
\altaffiltext{2}{Department of Astrophysical Sciences, Princeton University,
4 Ivy Lane, Princeton, NJ 08544, USA}
\altaffiltext{3}{Istituto Nazionale di Fisica Nucleare, Sezione Milano
Bicocca, gruppo collegato di Parma, I-43124 Parma, Italy}
\altaffiltext{4}{Dipartimento di Fisica, Universit\`{a} degli Studi di
Milano Bicocca, Piazza della Scienza 3, 20126 Milano, Italia}
\altaffiltext{5}{Dipartimento di Scienze Matematiche Fisiche ed
Informatiche, Universit\'a di Parma, I-43124 Parma, Italia}

\begin{abstract}
Gravitational waves detected from the binary neutron star (NS) merger
\GW{} constrained the NS equation of state by placing an upper bound
on certain parameters describing the binary's tidal interactions.
We show that the interpretation of the UV/optical/infrared counterpart
of \GW{} with kilonova models, combined with new numerical relativity
results, imply a complementary lower bound on the tidal deformability
parameter. The joint constraints tentatively rule out both extremely
stiff and soft NS equations of state.
\end{abstract}

\keywords{Gravitational waves -- Stars: neutron -- Equation of state}

\section{Introduction}

The properties of matter at supranuclear densities determining the
internal structure and mass-radius relation of \acp{NS}, are poorly
known at the moment \citep{ozel:2016oaf}. Presently, the strongest
constraint comes from the fact that the maximum mass for \acp{NS} must
be larger than about $2\ M_\odot$ \citep{antoniadis:2013pzd}.
Gravitational wave (GW) \acused{GW} observations of coalescing binary
\acp{NS} have long been considered as a promising avenue to constrain
the \ac{EOS} of dense matter. The tidal polarizability of the \acp{NS}
is encoded in the phase evolution of the \ac{GW} signal during the
inspiral \citep{flanagan:2007ix, hinderer:2009ca, damour:2009wj,
damour:2012yf, read:2013zra, delpozzo:2013ala, favata:2013rwa,
bernuzzi:2014owa, wade:2014vqa, lackey:2014fwa, hotokezaka:2016bzh,
hinderer:2016eia, lackey:2016krb, dietrich:2017aum, kiuchi:2017pte}. The
post-merger signal, if detected, could also place strong constraints on
the physics of high-density matter \citep{bauswein:2011tp,
takami:2014zpa, bernuzzi:2015rla, radice:2016rys, yang:2017xlf,
chatziioannou:2017ixj}.

On August 17, 2017, \acp{GW} from a pair of merging \acp{NS} were
observed, for the first time, by the LIGO-Virgo detector network
\citep{theligoscientific:2017qsa}: \GW. Less than 2 seconds after the
end of the \ac{GW} signal, a short $\gamma$-ray burst was detected by
the Fermi and INTEGRAL satellites in a coincident sky position
\citep{monitor:2017mdv}. In the following hours and days, the same
source, now named \AT, was detected in the X-ray, UV, optical, infrared,
and radio bands \citep{gbm:2017lvd, arcavi:2017a, chornock:2017sdf,
cowperthwaite:2017dyu, coulter:2017wya, drout:2017ijr, evans:2017mmy,
hallinan:2017woc, kasliwal:2017ngb, murguia-berthier:2017kkn,
nicholl:2017ahq, smartt:2017fuw, soares-santos:2017lru, tanvir:2017pws,
tanaka:2017qxj, troja:2017nqp}.

The preliminary analysis of \GW{} presented in
\citet{theligoscientific:2017qsa} already provided a first constraint on
the amplitude of tidal effects during the binary inspiral, disfavoring
\acp{EOS} with large \ac{NS} radii. \citet{margalit:2017dij} argued that
the merger remnant might not have formed a long lived remnant, because
of the relatively low energy of the ejecta inferred from optical and
infrared data. Under this assumption, \citet{margalit:2017dij}, and
subsequently \citet{shibata:2017xdx}, \citet{rezzolla:2017aly}, and
\citet{ruiz:2017due}, placed upper bounds on the maximum mass supported
by the \ac{NS} \ac{EOS}.  \citet{bauswein:2017vtn} pointed out that a
prompt \ac{BH} formation is also unlikely, because this would have
suppressed the ejection of matter and the subsequent emissions in the
optical/infrared. \citet{bauswein:2017vtn} combined this observation
with empirical relations between \ac{NS} radii and the threshold mass
for prompt collapse, which was previously found by means of simulations
with an approximate treatment of \ac{GR} \citep{bauswein:2013jpa}, to
tentatively rule out \acp{EOS} predicting very small \ac{NS} radii.

In this \emph{Letter} we propose and apply to \GW{} a new approach that
combines optical/infrared and \ac{GW} observations, by means of new
numerical relativity results, to derive strong joint constraints on the
tidal deformability of \acp{NS}.

\section{Multimessenger Observations}

The \ac{GW} data tightly constrained the 90\% credible interval for the
chirp mass of the binary, $\mathcal{M}_{\rm chirp} = (M_A M_B)^{3/5}
(M_A + M_B)^{-1/5}$, $M_A$ and $M_B$ being the \ac{NS} masses, to be
$1.188^{+0.004}_{-0.002}\ M_\odot$ \citep{theligoscientific:2017qsa}.
With the same confidence, the binary mass ratio $q = M_B/M_A$ is
constrained to be $0.7{-}1.0$ if the dimensionless \acp{NS} spins are
less than $0.05$ \citep{theligoscientific:2017qsa}.  If the priors on
the \ac{NS} spins are relaxed, $q$ becomes only constrained to be within
$0.4{-}1.0$. Note, however, that large spins are not expected on the
basis of the observed galactic \ac{NS} binary population
\citep{theligoscientific:2017qsa}. Moreover, $q < 0.7$ for this event
would imply an implausible mass for the secondary \ac{NS}, smaller than
$1.15\, M_\odot$, in tension with core-collapse supernova theory
\citep[e.g.,][]{radice:2017ykv}. Finally, we remark that the \ac{GW}
data already places strong limits on the component of the \ac{NS} spin
aligned with the orbital angular momentum
\citep{theligoscientific:2017qsa}.

LIGO and Virgo observations also constrain tidal effects in the inspiral
by placing an upper bound on the dimensionless quantity
\citep{flanagan:2007ix, favata:2013rwa}
\begin{equation}\label{eq:bns.lambda}
  \tilde{\Lambda} = \frac{16}{13} \left[\frac{(M_A + 12 M_B) M_A^4
  \tilde\Lambda_A }{(M_A + M_B)^5} + (A \leftrightarrow B)\right],
\end{equation}
which is inferred to be smaller than $800$ at the 90\% confidence level
\citep{theligoscientific:2017qsa}. In the previous equation
\begin{equation}\label{eq:lambda}
  \tilde\Lambda_i = \frac{2}{3} k_2^{(i)} \left[
    \left(\frac{c^2}{G}\right)\left(\frac{R_i}{M_i}\right)\right]^5\,,
  \quad i = A,B
\end{equation}
are the dimensionless quadrupolar tidal parameters (or tidal
polarizability coefficients), where $k_2^{(i)}$ are the quadrupolar Love
numbers for each star. The fate of the merger remnant is not known.
The postmerger high-frequency \acp{GW} were too weak to be detected, so
information on the remnant is not available from \ac{GW} observations
\citep{abbott:2017dke}.

The optical and infrared \ac{EM} data is well explained by the
radioactive decay of ${\sim}0.05\ M_\odot$ of material
\citep{chornock:2017sdf, cowperthwaite:2017dyu, drout:2017ijr,
nicholl:2017ahq, rosswog:2017sdn, tanaka:2017qxj, tanvir:2017pws,
perego:2017wtu, villar:2017wcc}. UV/optical light curve modeling of the
early emissions, hours to days after merger, points to the presence of a
relatively fast, $v\simeq 0.3\,c$, $M\simeq 0.02\ M_\odot$, component of
the outflow \citep{cowperthwaite:2017dyu, drout:2017ijr,
nicholl:2017ahq, perego:2017wtu, villar:2017wcc}. The modeling of the
later optical/infrared data points to the presence of at least another
component of the outflow with $v\simeq 0.1\,c$ and $M\simeq 0.04\
M_\odot$ \citep{chornock:2017sdf, cowperthwaite:2017dyu, drout:2017ijr,
perego:2017wtu, villar:2017wcc}. The inferred effective opacities for
these two (or more) outflow components suggest that they had different
compositions and, possibly, different origins.

\ac{GR} simulations indicate that only up to ${\sim}0.01\ M_\odot$ of
material can be unbound dynamically during the merger itself
\citep{hotokezaka:2012ze, bauswein:2013yna, radice:2016dwd,
lehner:2016lxy, sekiguchi:2016bjd, dietrich:2016lyp, bovard:2017mvn},
although larger ejecta masses can be reached for small mass ratios $q
\lesssim 0.6$ \citep{dietrich:2016hky}. The largest ejecta masses are
obtained for soft \acp{EOS}. In these cases, the outflows are fast, $v
\simeq (0.2{-}0.4)\,c$, shock heated, and re-processed by neutrinos
\citep{sekiguchi:2015dma, radice:2016dwd, foucart:2016rxm}.
Consequently, the dynamic ejecta can potentially explain the UV/optical
emissions in the first hours to days. The inferred properties for the
outflow component powering the optical/infrared emission on a days to
weeks timescale are more easily explained by neutrino, viscous, or
magnetically driven outflows from the merger remnant
\citep{dessart:2008zd, metzger:2008av, metzger:2008jt, fernandez:2013tya,
siegel:2014ita, just:2014fka, metzger:2014ila, perego:2014fma,
wu:2016pnw, siegel:2017nub, lippuner:2017bfm}. Detailed modeling
suggests that a disk mass of at least $0.08\, M_\odot$ is required to
explain \AT{} \citep{perego:2017wtu}.

\section{Simulation Results}

\begin{table*}
\caption{Gravitational and baryonic masses, compactnesses, tidal deformability
parameters, BH formation time, disk and ejecta masses. Disk and ejecta masses
are given at the final simulation time.}
\label{tab:summary}
\vspace{-1em}
\begin{center}
\begin{tabular}{ld{1.3}d{1.3}d{1.3}d{1.3}d{1.3}d{1.3}d{4.0}d{4.0}d{4.0}d{2.2}d{1.2}d{2.2}d{2.2}}
\toprule
EOS &
\myhead{$M_A$\tablenotemark{a}} &
\myhead{$M_B$\tablenotemark{a}} &
\myhead{$M_A^\ast$\tablenotemark{b}} &
\myhead{$M_B^\ast$\tablenotemark{b}} &
\myhead{$C_A$\tablenotemark{c}} &
\myhead{$C_B$\tablenotemark{c}} &
\myhead{$\tilde\Lambda_{A}$\tablenotemark{d}} &
\myhead{$\tilde\Lambda_{B}$\tablenotemark{d}} &
\myhead{$\tilde\Lambda$\tablenotemark{e}} &
\myhead{$M_{\rm disk}$\tablenotemark{f}} &
\myhead{$M_{\rm ej}$\tablenotemark{g}} &
\myhead{$t_{\rm BH}$\tablenotemark{h}} &
\myhead{$t_{\rm end}$\tablenotemark{i}} \\
&
\myhead{$[M_\odot]$} &
\myhead{$[M_\odot]$} &
\myhead{$[M_\odot]$} &
\myhead{$[M_\odot]$} &
& & & & &
\multicolumn{2}{c}{$[10^{-2}\ M_\odot]$} &
\myhead{$[{\rm ms}]$} &
\myhead{$[{\rm ms}]$} \\
\midrule
BHB$\Lambda\phi$ & 1.365 & 1.25 & 1.491 & 1.352 & 0.153 & 0.140 &  805 & 1310 &1028 & 18.73 & 0.06 & \myhead{$-$} & 23.98 \\
BHB$\Lambda\phi$ & 1.35 & 1.35 & 1.473 & 1.473 & 0.151 & 0.151 &  857 &  857 & 857 & 14.45 & 0.07 & \myhead{$-$} & 21.26 \\
BHB$\Lambda\phi$ & 1.4 & 1.2 & 1.533 & 1.297 & 0.157 & 0.135 &  697 & 1630 &1068 & 20.74 & 0.11 & \myhead{$-$} & 23.74 \\
BHB$\Lambda\phi$ & 1.4 & 1.4 & 1.533 & 1.533 & 0.157 & 0.157 &  697 &  697 & 697 & 7.05 & 0.09 & 11.96 & 16.39 \\
BHB$\Lambda\phi$ & 1.44 & 1.39 & 1.580 & 1.520 & 0.161 & 0.155 &  591 &  726 & 655 & 8.28 & 0.06 & 10.39 & 15.77 \\
BHB$\Lambda\phi$ & 1.5 & 1.5 & 1.657 & 1.657 & 0.168 & 0.168 &  462 &  462 & 462 & 1.93 & 0.05 & 2.27 & 11.78 \\
BHB$\Lambda\phi$ & 1.6 & 1.6 & 1.778 & 1.778 & 0.179 & 0.179 &  306 &  306 & 306 & 0.09 & 0.00 & 0.99 & 10.67 \\
DD2 & 1.365 & 1.25 & 1.491 & 1.352 & 0.153 & 0.140 &  807 & 1309 &1028 & 20.83 & 0.04 & \myhead{$-$} & 24.24 \\
DD2 & 1.35 & 1.35 & 1.473 & 1.473 & 0.151 & 0.151 &  858 &  858 & 858 & 15.69 & 0.03 & \myhead{$-$} & 24.41 \\
DD2 & 1.4 & 1.2 & 1.533 & 1.297 & 0.157 & 0.135 &  699 & 1630 &1070 & 19.26 & 0.09 & \myhead{$-$} & 23.59 \\
DD2 & 1.4 & 1.4 & 1.533 & 1.533 & 0.157 & 0.157 &  699 &  699 & 699 & 12.36 & 0.04 & \myhead{$-$} & 24.52 \\
DD2 & 1.44 & 1.39 & 1.580 & 1.520 & 0.161 & 0.155 &  595 &  728 & 658 & 14.40 & 0.05 & \myhead{$-$} & 23.52 \\
DD2 & 1.5 & 1.5 & 1.657 & 1.657 & 0.167 & 0.167 &  469 &  469 & 469 & 16.70 & 0.07 & \myhead{$-$} & 23.12 \\
DD2 & 1.6 & 1.6 & 1.778 & 1.778 & 0.178 & 0.178 &  317 &  317 & 317 & 1.96 & 0.12 & 2.28 & 12.08 \\
LS220 & 1.2 & 1.2 & 1.296 & 1.296 & 0.139 & 0.139 & 1439 & 1439 &1439 & 17.43 & 0.14 & \myhead{$-$} & 23.22 \\
LS220 & 1.365 & 1.25 & 1.491 & 1.355 & 0.159 & 0.145 &  636 & 1119 & 848 & 16.86 & 0.11 & \myhead{$-$} & 26.71 \\
LS220 & 1.35 & 1.35 & 1.473 & 1.473 & 0.157 & 0.157 &  684 &  684 & 684 & 7.25 & 0.06 & 20.34 & 23.84 \\
LS220 & 1.4 & 1.2 & 1.535 & 1.296 & 0.163 & 0.139 &  536 & 1439 & 893 & 22.82 & 0.19 & \myhead{$-$} & 23.52 \\
LS220 & 1.4 & 1.4 & 1.535 & 1.535 & 0.163 & 0.163 &  536 &  536 & 536 & 4.58 & 0.14 & 9.93 & 26.95 \\
LS220 & 1.44 & 1.39 & 1.581 & 1.520 & 0.168 & 0.162 &  442 &  563 & 499 & 3.91 & 0.19 & 7.22 & 14.83 \\
LS220 & 1.45 & 1.45 & 1.596 & 1.596 & 0.169 & 0.169 &  421 &  421 & 421 & 2.05 & 0.16 & 2.26 & 11.83 \\
LS220 & 1.6 & 1.6 & 1.790 & 1.790 & 0.189 & 0.189 &  202 &  202 & 202 & 0.07 & 0.03 & 0.63 & 10.42 \\
LS220 & 1.71 & 1.71 & 1.928 & 1.928 & 0.205 & 0.205 &  116 &  116 & 116 & 0.06 & 0.03 & 0.49 & 9.94 \\
SFHo & 1.365 & 1.25 & 1.504 & 1.364 & 0.169 & 0.155 &  393 &  680 & 520 & 8.81 & 0.15 & \myhead{$-$} & 26.41 \\
SFHo & 1.35 & 1.35 & 1.486 & 1.486 & 0.167 & 0.167 &  422 &  422 & 422 & 6.23 & 0.35 & 11.96 & 22.88 \\
SFHo & 1.4 & 1.2 & 1.547 & 1.303 & 0.174 & 0.148 &  334 &  868 & 546 & 11.73 & 0.12 & \myhead{$-$} & 24.31 \\
SFHo & 1.4 & 1.4 & 1.547 & 1.547 & 0.174 & 0.174 &  334 &  334 & 334 & 0.01 & 0.04 & 1.07 & 13.91 \\
SFHo & 1.44 & 1.39 & 1.598 & 1.535 & 0.179 & 0.173 &  277 &  350 & 312 & 0.09 & 0.04 & 0.87 & 7.06 \\
SFHo & 1.46 & 1.46 & 1.623 & 1.623 & 0.182 & 0.182 &  252 &  252 & 252 & 0.02 & 0.00 & 0.70 & 9.51 \\
\bottomrule\vspace{-1em}
\end{tabular}
\vspace{-0.5em}
\tablenotetext{1}{NS gravitational mass.}
\tablenotetext{2}{NS baryonic mass.}
\tablenotetext{3}{NS compactness, $G M / R\, c^2$.}
\tablenotetext{4}{Dimensionless quadrupolar tidal parameters, Eq.~(\ref{eq:lambda}).}
\tablenotetext{5}{Dimensionless tidal parameter, Eq.~(\ref{eq:bns.lambda}).}
\tablenotetext{6}{Gravitationally bound material with $\rho \leq
10^{13}\ {\rm g}\ {\rm cm}^{-3}$ outside of the apparent horizon.}
\tablenotetext{7}{Dynamic ejecta mass, computed as from the flux of
unbound matter through the coordinate-sphere $r = 443\ {\rm km}$.}
\tablenotetext{8}{BH formation time, in milliseconds after merger.}
\tablenotetext{9}{Final simulation time, in milliseconds after merger.}
\end{center}
\end{table*}

We perform 29 merger simulations using the GR hydrodynamics code
\texttt{WhiskyTHC} \citep{radice:2012cu, radice:2013hxh,
radice:2013xpa}. We consider both equal and unequal mass configurations,
and we adopt 4 temperature and composition dependent nuclear \acp{EOS}
spanning the range of the nuclear uncertainties: the DD2 EOS
\citep{typel:2009sy, hempel:2009mc}, the BHB$\Lambda\phi$ EOS
\citep{banik:2014qja}, the LS220 EOS \citep{lattimer:1991nc}, and the
SFHo EOS \citep{steiner:2012rk}. This is the largest dataset of
simulations performed in full-GR and with realistic microphysics to
date. Neutrino cooling and $Y_e$ evolution are treated as discussed in
\citet{radice:2016dwd}. The computational setup is the same as in
\citet{radice:2016rys}. The resolution of the grid regions covering the
\acp{NS} and the merger remnant is $\simeq 185\ {\rm m}$. We verify the
robustness of our results and estimate the numerical uncertainties by
performing 6 additional simulations at $25\%$ higher resolution.
We conservatively estimate finite-resolution error on the disk
and dynamic ejecta masses to be
\begin{equation}\label{eq:errors}
  \Delta M_{\rm disk, ej} = 0.5\, M_{\rm disk, ej} + \epsilon_{\rm disk, ej}\,,
\end{equation}
where $\epsilon_{\rm disk} = 5 \times 10^{-4} M_\odot$ and
$\epsilon_{\rm ej} = 5 \times 10^{-5} M_\odot$.  A more detailed account
of these simulations will be given elsewhere (Radice et al., in
prep.~2017). A summary of the simulations is given in
Tab.~\ref{tab:summary}.

\begin{figure}
  \includegraphics[width=\columnwidth]{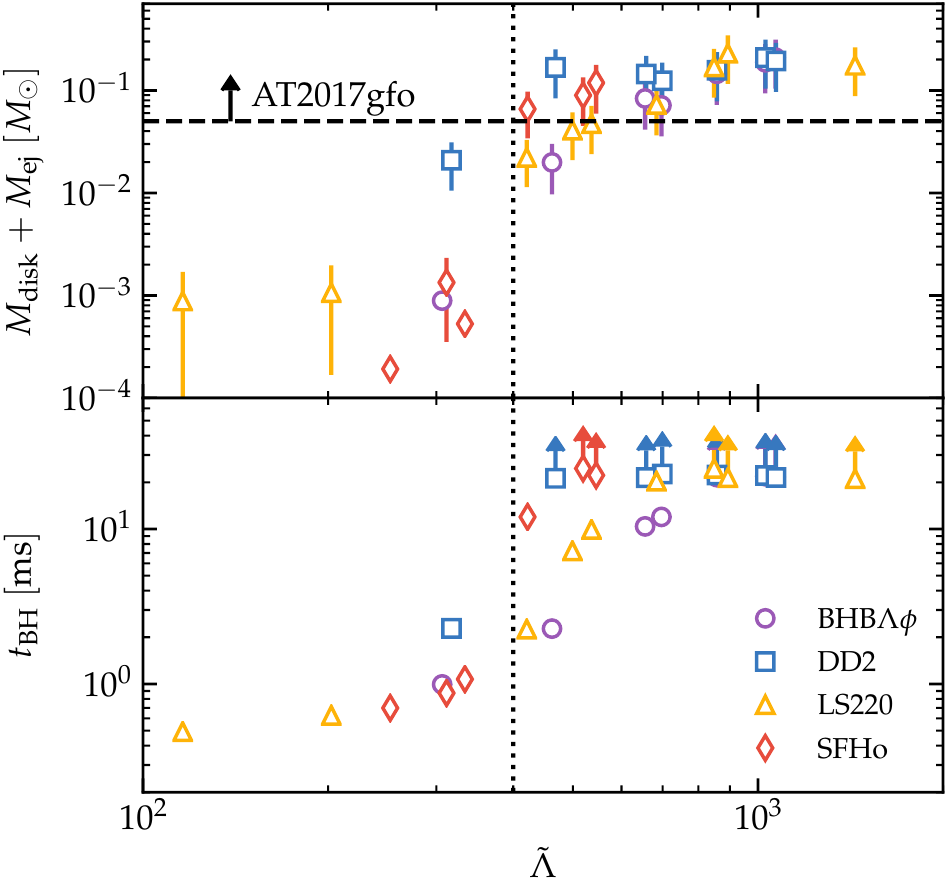}
  \caption{Remnant disk plus dynamic ejecta masses (\emph{upper panel})
  and BH formation time (\emph{lower panel}) plotted against the tidal
  parameter $\tilde\Lambda$ (Eq.~\ref{eq:bns.lambda}). For models that
  do not collapse during our simulation time, we give a lower limit.
  The horizontal dashed line shows a conservative lower limit for \AT{},
  $0.05\, M_\odot$, obtained assuming that the entire disk is unbound.
  The vertical dotted line is $\tilde\Lambda = 400$. Errors on
  $M_{\rm disk}$ and $M_{\rm ej}$ are estimated following
  Eq.~(\ref{eq:errors}) and are added in quadrature.}
  \label{fig:remnants}
\end{figure}

We compute the mass of the dynamic ejecta and of the remnant accretion
disk for each model. Our results are shown in Tab.~\ref{tab:summary} and
Fig.~\ref{fig:remnants}. The typical dynamic ejecta mass in our
simulations are of the order of ${\sim}10^{-3}\ M_\odot$, in good
qualitative agreement with previous numerical relativity results. We do
not find any clear indication of a trend in the dynamic ejecta masses as
a function of the binary parameters or \ac{EOS}.  However, we find a
clear correlation between the disk masses and the tidal parameter
$\tilde\Lambda$. According to our simulations, binaries with
$\tilde\Lambda \lesssim 450$ inevitably produce \acp{BH} with small
$\lesssim 10^{-2}\, M_\odot$ accretion disks. These cases are
incompatible with the infrared data for \AT{}, even under the assumption
that all of the matter left outside of the event horizon will be
ejected.

The reason for this trend is easily understood from the lower panel of
Fig.~\ref{fig:remnants}. The \ac{NS} dimensionless quadrupolar tidal
parameters depend on the negative-fifth power of the \ac{NS} compactness
($G M/R\, c^2$; Eq.~\ref{eq:lambda}). Consequently, small values of
$\tilde\Lambda$ are associated with binary systems having compact
\acp{NS} that result in rapid or prompt \ac{BH} formation. In these
cases, the collapse happens on a shorter timescale than the hydrodynamic
processes responsible for the formation of the disk. Consequently, only
a small amount of mass is left outside of the event horizon at the
end of the simulations.

Binaries with larger values of $\tilde\Lambda$ produce more massive
disks, up to ${\sim}0.2\ M_\odot$, and longer lived remnants. In these
cases, neutrino driven winds and viscous and magnetic processes in the
disk are expected to unbind sufficient material to explain the optical
and infrared observations for \AT{} \citep{perego:2014fma, wu:2016pnw,
siegel:2017nub}.

\section{Discussion}

\begin{figure}
  \includegraphics[width=\columnwidth]{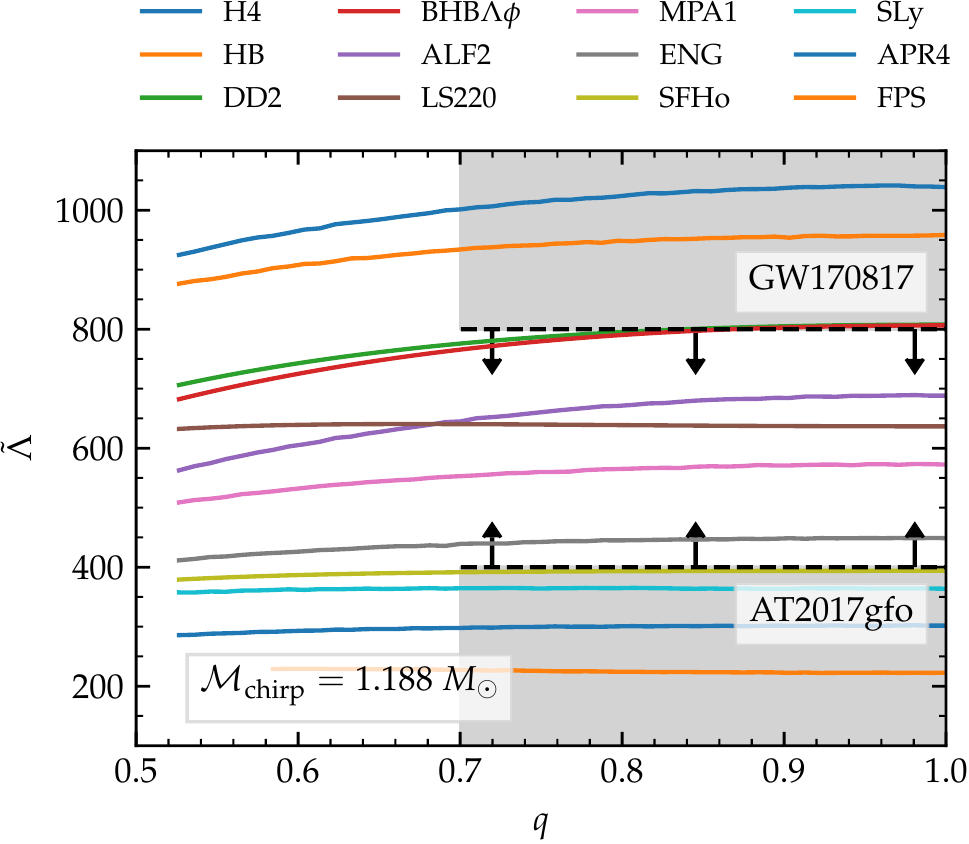}
  \caption{Tidal parameter $\tilde\Lambda$ (Eq.~\ref{eq:bns.lambda}) as
  a function of the mass ratio $q$ for a fixed chirp mass
  $\mathcal{M}_{\rm chirp} = 1.188\ M_\odot$. The shaded region shows
  the region excluded with 90\% confidence level by the LIGO-Virgo
  observations \citep{theligoscientific:2017qsa}, with the additional
  constraint of $\tilde\Lambda \geq 400$ derived from the simulations
  and the EM observations. EOSs whose curves enter this region are
  disfavored. EOSs are sorted for decreasing $\tilde\Lambda$ at $q=1$,
  i.e., H4 is the stiffest EOS in our sample, and FPS is the softest.}
  \label{fig:lambdas}
\end{figure}

On the basis of our simulations and the current interpretation of the
UV/optical/infrared data we can conclude that values of $\tilde\Lambda$
smaller than $400$ are tentatively excluded. Together with the
LIGO-Virgo constraints on $\tilde\Lambda$
\citep{theligoscientific:2017qsa}, this result already yields a strong
constraint on the \ac{EOS}.

To illustrate this, we notice that, since the chirp mass of the binary
progenitor of \GW{} is well measured, for any given \ac{EOS} the
predicted $\tilde\Lambda$ reduces to a simple function of the mass
ratio, that is,
\begin{equation}
  \tilde\Lambda = \tilde\Lambda\left(q, \mathcal{M}_{\rm
  chirp} = 1.188\, M_\odot; \mathrm{EOS}\right).
\end{equation}
We consider a set of 12 \acp{EOS}: the four used in the simulations and
other eight from \citet{read:2008iy}. We compute $\tilde\Lambda(q)$ for
each and show the resulting curves in Fig.~\ref{fig:lambdas}. There, we
also show the upper bound on $\tilde\Lambda$ from the \ac{GW}
observations as well as the newly estimated lower bound from the \ac{EM}
data. On the one hand, stiff \acp{EOS}, such as H4 and HB, are already
disfavored on the basis of the \ac{GW} data alone. On the other hand,
\ac{EOS} as soft as FPS and APR4 are also tentatively excluded on the
basis of the \ac{EM} observations\footnote{Note that FPS is also
excluded because it predicts a maximum \ac{NS} mass smaller than $2\
M_\odot$.}. Soft \ac{EOS} commonly used in simulations, such as SFHo
and SLy, lay at the lower boundary of the allowed region, while DD2
and BHB$\Lambda\phi$ are on the upper boundary.

Our results show that numerical relativity simulations are key to
exploiting the potential of multimessenger observations. While \ac{GW}
data bounds the tidal deformability of \acp{NS} from above, the \ac{EM}
data and our simulations bound it from below. The result is a
competitive constraint already after the first detection of a merger
event. Our method is general, it can be applied to future observations
and used to inform the priors used in the \ac{GW} data analysis. We
anticipate that, with more observations and more precise simulations,
the bounds on the tidal deformability of NSs will be further improved.

The physics setting the lower bound on $\tilde\Lambda$ is well
understood and under control in our simulations. However, a more
extended analysis taking into account the uncertainties in the
interpretation of the \ac{EM} observations and in the simulations is a
necessary next step. For example, large components of the \ac{NS} spins
parallel to the orbital plane are not expected, but also not constrained
for \GW{}. We cannot exclude that, if present, they will affect our
results.
Moreover, there are indication that small mass ratio binaries $q
\lesssim 0.8$ might also form disks with masses up to ${\sim}0.1\,
M_\odot$ \citep{shibata:2017xdx}. If confirmed, this would imply that
the lower bound on $\tilde\Lambda$ might depend on $q$.  
Note that the upper-bound on $\tilde\Lambda$ estimated from the
\ac{GW} signal is also likely to have some dependency on $q$.
Consequently, a more precise determination of the exclusion region on
$\tilde\Lambda$ will necessarily require a full Bayesian analysis of the
\ac{GW} data using $\tilde\Lambda$ priors informed by
numerical-relativity results.
We plan to improve our modeling by means of new simulations
exploring the set of binary progenitor parameters compatible with \GW{}
and the associated \ac{EM} counterparts.

\begin{acknowledgments}
It is a pleasure to acknowledge A.~Burrows for the many stimulating
discussions, and  T.~Venumadhav for comments on an earlier version of
the manuscript.
DR acknowledges support from a Frank and Peggy Taplin Membership at the
Institute for Advanced Study and the Max-Planck/Princeton Center (MPPC)
for Plasma Physics (NSF PHY-1523261).
DR and AP acknowledge support from the Institute for Nuclear Theory
(17-2b program).
SB acknowledges support by the EU H2020 under ERC Starting Grant,
no.~BinGraSp-714626.
Computations were performed on the supercomputers Bridges, Comet, and
Stampede (NSF XSEDE allocation TG-PHY160025), on NSF/NCSA Blue Waters
(NSF PRAC ACI-1440083), Marconi (PRACE proposal 2016153522), and
PizDaint/CSCS (ID 667).
This manuscript has been assigned LIGO report number LIGO-P1700421 and
Virgo report number VIR-0894A-17.
\end{acknowledgments}


\begin{thebibliography}{82}
\expandafter\ifx\csname natexlab\endcsname\relax\def\natexlab#1{#1}\fi

\bibitem[{Abbott {et~al.}(2017{\natexlab{a}})}]{monitor:2017mdv}
Abbott, B.~P., {et~al.} 2017{\natexlab{a}}, Astrophys. J., 848, L13, 1710.05834

\bibitem[{Abbott {et~al.}(2017{\natexlab{b}})}]{theligoscientific:2017qsa}
------. 2017{\natexlab{b}}, Phys. Rev. Lett., 119, 161101, 1710.05832

\bibitem[{Abbott {et~al.}(2017{\natexlab{c}})}]{gbm:2017lvd}
------. 2017{\natexlab{c}}, Astrophys. J., 848, L12, 1710.05833

\bibitem[{Abbott {et~al.}(2017{\natexlab{d}})}]{abbott:2017dke}
------. 2017{\natexlab{d}}, Astrophys. J., 851, L16, 1710.09320

\bibitem[{Antoniadis {et~al.}(2013)}]{antoniadis:2013pzd}
Antoniadis, J., {et~al.} 2013, Science, 340, 6131, 1304.6875

\bibitem[{Arcavi {et~al.}(2017)Arcavi, McCully, Hosseinzadeh, Howell, Vasylyev,
  Poznanski, Zaltzman, Maoz, Singer, Valenti, Kasen, Barnes, Piran, \& fai
  Fong}]{arcavi:2017a}
Arcavi, I. {et~al.} 2017, The Astrophysical Journal, 848, L33

\bibitem[{Banik {et~al.}(2014)Banik, Hempel, \& Bandyopadhyay}]{banik:2014qja}
Banik, S., Hempel, M., \& Bandyopadhyay, D. 2014, Astrophys. J. Suppl., 214,
  22, 1404.6173

\bibitem[{Bauswein {et~al.}(2013{\natexlab{a}})Bauswein, Baumgarte, \&
  Janka}]{bauswein:2013jpa}
Bauswein, A., Baumgarte, T.~W., \& Janka, H.~T. 2013{\natexlab{a}}, Phys. Rev.
  Lett., 111, 131101, 1307.5191

\bibitem[{Bauswein {et~al.}(2013{\natexlab{b}})Bauswein, Goriely, \&
  Janka}]{bauswein:2013yna}
Bauswein, A., Goriely, S., \& Janka, H.~T. 2013{\natexlab{b}}, Astrophys. J.,
  773, 78, 1302.6530

\bibitem[{Bauswein \& Janka(2012)}]{bauswein:2011tp}
Bauswein, A., \& Janka, H.~T. 2012, Phys. Rev. Lett., 108, 011101, 1106.1616

\bibitem[{Bauswein {et~al.}(2017)Bauswein, Just, Janka, \&
  Stergioulas}]{bauswein:2017vtn}
Bauswein, A., Just, O., Janka, H.-T., \& Stergioulas, N. 2017, Astrophys. J.,
  850, L34, 1710.06843

\bibitem[{Bernuzzi {et~al.}(2015{\natexlab{a}})Bernuzzi, Dietrich, \&
  Nagar}]{bernuzzi:2015rla}
Bernuzzi, S., Dietrich, T., \& Nagar, A. 2015{\natexlab{a}}, Phys. Rev. Lett.,
  115, 091101, 1504.01764

\bibitem[{Bernuzzi {et~al.}(2015{\natexlab{b}})Bernuzzi, Nagar, Dietrich, \&
  Damour}]{bernuzzi:2014owa}
Bernuzzi, S., Nagar, A., Dietrich, T., \& Damour, T. 2015{\natexlab{b}}, Phys.
  Rev. Lett., 114, 161103, 1412.4553

\bibitem[{Bovard {et~al.}(2017)Bovard, Martin, Guercilena, Arcones, Rezzolla,
  \& Korobkin}]{bovard:2017mvn}
Bovard, L., Martin, D., Guercilena, F., Arcones, A., Rezzolla, L., \& Korobkin,
  O. 2017, Phys. Rev., D96, 124005, 1709.09630

\bibitem[{Chatziioannou {et~al.}(2017)Chatziioannou, Clark, Bauswein,
  Millhouse, Littenberg, \& Cornish}]{chatziioannou:2017ixj}
Chatziioannou, K., Clark, J.~A., Bauswein, A., Millhouse, M., Littenberg,
  T.~B., \& Cornish, N. 2017, 1711.00040

\bibitem[{Chornock {et~al.}(2017)}]{chornock:2017sdf}
Chornock, R., {et~al.} 2017, Astrophys. J., 848, L19, 1710.05454

\bibitem[{Coulter {et~al.}(2017)}]{coulter:2017wya}
Coulter, D.~A., {et~al.} 2017, Science, 1710.05452

\bibitem[{Cowperthwaite {et~al.}(2017)}]{cowperthwaite:2017dyu}
Cowperthwaite, P.~S., {et~al.} 2017, Astrophys. J., 848, L17, 1710.05840

\bibitem[{Damour \& Nagar(2010)}]{damour:2009wj}
Damour, T., \& Nagar, A. 2010, Phys. Rev., D81, 084016, 0911.5041

\bibitem[{Damour {et~al.}(2012)Damour, Nagar, \& Villain}]{damour:2012yf}
Damour, T., Nagar, A., \& Villain, L. 2012, Phys. Rev., D85, 123007, 1203.4352

\bibitem[{Del~Pozzo {et~al.}(2013)Del~Pozzo, Li, Agathos, Van Den~Broeck, \&
  Vitale}]{delpozzo:2013ala}
Del~Pozzo, W., Li, T. G.~F., Agathos, M., Van Den~Broeck, C., \& Vitale, S.
  2013, Phys. Rev. Lett., 111, 071101, 1307.8338

\bibitem[{Dessart {et~al.}(2009)Dessart, Ott, Burrows, Rosswog, \&
  Livne}]{dessart:2008zd}
Dessart, L., Ott, C., Burrows, A., Rosswog, S., \& Livne, E. 2009, Astrophys.
  J., 690, 1681, 0806.4380

\bibitem[{Dietrich {et~al.}(2017{\natexlab{a}})Dietrich, Bernuzzi, \&
  Tichy}]{dietrich:2017aum}
Dietrich, T., Bernuzzi, S., \& Tichy, W. 2017{\natexlab{a}}, 1706.02969

\bibitem[{Dietrich {et~al.}(2017{\natexlab{b}})Dietrich, Bernuzzi, Ujevic, \&
  Tichy}]{dietrich:2016lyp}
Dietrich, T., Bernuzzi, S., Ujevic, M., \& Tichy, W. 2017{\natexlab{b}}, Phys.
  Rev., D95, 044045, 1611.07367

\bibitem[{Dietrich {et~al.}(2017{\natexlab{c}})Dietrich, Ujevic, Tichy,
  Bernuzzi, \& Bruegmann}]{dietrich:2016hky}
Dietrich, T., Ujevic, M., Tichy, W., Bernuzzi, S., \& Bruegmann, B.
  2017{\natexlab{c}}, Phys. Rev., D95, 024029, 1607.06636

\bibitem[{Drout {et~al.}(2017)}]{drout:2017ijr}
Drout, M.~R., {et~al.} 2017, Science, 1710.05443

\bibitem[{Evans {et~al.}(2017)}]{evans:2017mmy}
Evans, P.~A., {et~al.} 2017, Science, 1710.05437

\bibitem[{Favata(2014)}]{favata:2013rwa}
Favata, M. 2014, Phys. Rev. Lett., 112, 101101, 1310.8288

\bibitem[{Fernández \& Metzger(2013)}]{fernandez:2013tya}
Fernández, R., \& Metzger, B.~D. 2013, Mon. Not. Roy. Astron. Soc., 435, 502,
  1304.6720

\bibitem[{Flanagan \& Hinderer(2008)}]{flanagan:2007ix}
Flanagan, E.~E., \& Hinderer, T. 2008, Phys. Rev., D77, 021502, 0709.1915

\bibitem[{Foucart {et~al.}(2016)Foucart, O'Connor, Roberts, Kidder, Pfeiffer,
  \& Scheel}]{foucart:2016rxm}
Foucart, F., O'Connor, E., Roberts, L., Kidder, L.~E., Pfeiffer, H.~P., \&
  Scheel, M.~A. 2016, Phys. Rev., D94, 123016, 1607.07450

\bibitem[{Hallinan {et~al.}(2017)}]{hallinan:2017woc}
Hallinan, G., {et~al.} 2017, Science, 1710.05435

\bibitem[{Hempel \& Schaffner-Bielich(2010)}]{hempel:2009mc}
Hempel, M., \& Schaffner-Bielich, J. 2010, Nucl. Phys., A837, 210, 0911.4073

\bibitem[{Hinderer {et~al.}(2010)Hinderer, Lackey, Lang, \&
  Read}]{hinderer:2009ca}
Hinderer, T., Lackey, B.~D., Lang, R.~N., \& Read, J.~S. 2010, Phys. Rev., D81,
  123016, 0911.3535

\bibitem[{Hinderer {et~al.}(2016)}]{hinderer:2016eia}
Hinderer, T., {et~al.} 2016, Phys. Rev. Lett., 116, 181101, 1602.00599

\bibitem[{Hotokezaka {et~al.}(2013)Hotokezaka, Kiuchi, Kyutoku, Okawa,
  Sekiguchi, Shibata, \& Taniguchi}]{hotokezaka:2012ze}
Hotokezaka, K., Kiuchi, K., Kyutoku, K., Okawa, H., Sekiguchi, Y.-i., Shibata,
  M., \& Taniguchi, K. 2013, Phys. Rev., D87, 024001, 1212.0905

\bibitem[{Hotokezaka {et~al.}(2016)Hotokezaka, Kyutoku, Sekiguchi, \&
  Shibata}]{hotokezaka:2016bzh}
Hotokezaka, K., Kyutoku, K., Sekiguchi, Y.-i., \& Shibata, M. 2016, Phys. Rev.,
  D93, 064082, 1603.01286

\bibitem[{Just {et~al.}(2015)Just, Bauswein, Pulpillo, Goriely, \&
  Janka}]{just:2014fka}
Just, O., Bauswein, A., Pulpillo, R.~A., Goriely, S., \& Janka, H.~T. 2015,
  Mon. Not. Roy. Astron. Soc., 448, 541, 1406.2687

\bibitem[{Kasliwal {et~al.}(2017)}]{kasliwal:2017ngb}
Kasliwal, M.~M., {et~al.} 2017, Science, 1710.05436

\bibitem[{Kiuchi {et~al.}(2017)Kiuchi, Kawaguchi, Kyutoku, Sekiguchi, Shibata,
  \& Taniguchi}]{kiuchi:2017pte}
Kiuchi, K., Kawaguchi, K., Kyutoku, K., Sekiguchi, Y., Shibata, M., \&
  Taniguchi, K. 2017, Phys. Rev., D96, 084060, 1708.08926

\bibitem[{Lackey {et~al.}(2016)Lackey, Bernuzzi, Galley, Meidam, \& Van
  Den~Broeck}]{lackey:2016krb}
Lackey, B.~D., Bernuzzi, S., Galley, C.~R., Meidam, J., \& Van Den~Broeck, C.
  2016, 1610.04742

\bibitem[{Lackey \& Wade(2015)}]{lackey:2014fwa}
Lackey, B.~D., \& Wade, L. 2015, Phys. Rev., D91, 043002, 1410.8866

\bibitem[{Lattimer \& Swesty(1991)}]{lattimer:1991nc}
Lattimer, J.~M., \& Swesty, F.~D. 1991, Nucl. Phys., A535, 331

\bibitem[{Lehner {et~al.}(2016)Lehner, Liebling, Palenzuela, Caballero,
  O'Connor, Anderson, \& Neilsen}]{lehner:2016lxy}
Lehner, L., Liebling, S.~L., Palenzuela, C., Caballero, O.~L., O'Connor, E.,
  Anderson, M., \& Neilsen, D. 2016, Class. Quant. Grav., 33, 184002,
  1603.00501

\bibitem[{Lippuner {et~al.}(2017)Lippuner, Fernández, Roberts, Foucart, Kasen,
  Metzger, \& Ott}]{lippuner:2017bfm}
Lippuner, J., Fernández, R., Roberts, L.~F., Foucart, F., Kasen, D., Metzger,
  B.~D., \& Ott, C.~D. 2017, Mon. Not. Roy. Astron. Soc., 472, 904, 1703.06216

\bibitem[{Margalit \& Metzger(2017)}]{margalit:2017dij}
Margalit, B., \& Metzger, B.~D. 2017, Astrophys. J., 850, L19, 1710.05938

\bibitem[{Metzger \& Fernández(2014)}]{metzger:2014ila}
Metzger, B.~D., \& Fernández, R. 2014, Mon. Not. Roy. Astron. Soc., 441, 3444,
  1402.4803

\bibitem[{Metzger {et~al.}(2008)Metzger, Piro, \& Quataert}]{metzger:2008av}
Metzger, B.~D., Piro, A.~L., \& Quataert, E. 2008, Mon. Not. Roy. Astron. Soc.,
  390, 781, 0805.4415

\bibitem[{Metzger {et~al.}(2009)Metzger, Piro, \& Quataert}]{metzger:2008jt}
------. 2009, Mon. Not. Roy. Astron. Soc., 396, 304, 0810.2535

\bibitem[{Murguia-Berthier {et~al.}(2017)}]{murguia-berthier:2017kkn}
Murguia-Berthier, A., {et~al.} 2017, 1710.05453

\bibitem[{Nicholl {et~al.}(2017)}]{nicholl:2017ahq}
Nicholl, M., {et~al.} 2017, Astrophys. J., 848, L18, 1710.05456

\bibitem[{Ozel \& Freire(2016)}]{ozel:2016oaf}
Ozel, F., \& Freire, P. 2016, Ann. Rev. Astron. Astrophys., 54, 401, 1603.02698

\bibitem[{Perego {et~al.}(2017)Perego, Radice, \& Bernuzzi}]{perego:2017wtu}
Perego, A., Radice, D., \& Bernuzzi, S. 2017, Astrophys. J., 850, L37,
  1711.03982

\bibitem[{Perego {et~al.}(2014)Perego, Rosswog, Cabezón, Korobkin, Käppeli,
  Arcones, \& Liebendörfer}]{perego:2014fma}
Perego, A., Rosswog, S., Cabezón, R.~M., Korobkin, O., Käppeli, R., Arcones,
  A., \& Liebendörfer, M. 2014, Mon. Not. Roy. Astron. Soc., 443, 3134,
  1405.6730

\bibitem[{Radice {et~al.}(2017{\natexlab{a}})Radice, Bernuzzi, Del~Pozzo,
  Roberts, \& Ott}]{radice:2016rys}
Radice, D., Bernuzzi, S., Del~Pozzo, W., Roberts, L.~F., \& Ott, C.~D.
  2017{\natexlab{a}}, Astrophys. J., 842, L10, 1612.06429

\bibitem[{Radice {et~al.}(2017{\natexlab{b}})Radice, Burrows, Vartanyan,
  Skinner, \& Dolence}]{radice:2017ykv}
Radice, D., Burrows, A., Vartanyan, D., Skinner, M.~A., \& Dolence, J.~C.
  2017{\natexlab{b}}, Astrophys. J., 850, 43, 1702.03927

\bibitem[{Radice {et~al.}(2016)Radice, Galeazzi, Lippuner, Roberts, Ott, \&
  Rezzolla}]{radice:2016dwd}
Radice, D., Galeazzi, F., Lippuner, J., Roberts, L.~F., Ott, C.~D., \&
  Rezzolla, L. 2016, Mon. Not. Roy. Astron. Soc., 460, 3255, 1601.02426

\bibitem[{Radice \& Rezzolla(2012)}]{radice:2012cu}
Radice, D., \& Rezzolla, L. 2012, Astron. Astrophys., 547, A26, 1206.6502

\bibitem[{Radice {et~al.}(2014{\natexlab{a}})Radice, Rezzolla, \&
  Galeazzi}]{radice:2013hxh}
Radice, D., Rezzolla, L., \& Galeazzi, F. 2014{\natexlab{a}}, Mon. Not. Roy.
  Astron. Soc., 437, L46, 1306.6052

\bibitem[{Radice {et~al.}(2014{\natexlab{b}})Radice, Rezzolla, \&
  Galeazzi}]{radice:2013xpa}
------. 2014{\natexlab{b}}, Class. Quant. Grav., 31, 075012, 1312.5004

\bibitem[{Read {et~al.}(2013)Read, Baiotti, Creighton, Friedman, Giacomazzo,
  Kyutoku, Markakis, Rezzolla, Shibata, \& Taniguchi}]{read:2013zra}
Read, J.~S. {et~al.} 2013, Phys. Rev., D88, 044042, 1306.4065

\bibitem[{Read {et~al.}(2009)Read, Lackey, Owen, \& Friedman}]{read:2008iy}
Read, J.~S., Lackey, B.~D., Owen, B.~J., \& Friedman, J.~L. 2009, Phys. Rev.,
  D79, 124032, 0812.2163

\bibitem[{Rezzolla {et~al.}(2017)Rezzolla, Most, \& Weih}]{rezzolla:2017aly}
Rezzolla, L., Most, E.~R., \& Weih, L.~R. 2017, 1711.00314

\bibitem[{Rosswog {et~al.}(2017)Rosswog, Sollerman, Feindt, Goobar, Korobkin,
  Fremling, \& Kasliwal}]{rosswog:2017sdn}
Rosswog, S., Sollerman, J., Feindt, U., Goobar, A., Korobkin, O., Fremling, C.,
  \& Kasliwal, M. 2017, 1710.05445

\bibitem[{Ruiz {et~al.}(2017)Ruiz, Shapiro, \& Tsokaros}]{ruiz:2017due}
Ruiz, M., Shapiro, S.~L., \& Tsokaros, A. 2017, 1711.00473

\bibitem[{Sekiguchi {et~al.}(2015)Sekiguchi, Kiuchi, Kyutoku, \&
  Shibata}]{sekiguchi:2015dma}
Sekiguchi, Y., Kiuchi, K., Kyutoku, K., \& Shibata, M. 2015, Phys. Rev., D91,
  064059, 1502.06660

\bibitem[{Sekiguchi {et~al.}(2016)Sekiguchi, Kiuchi, Kyutoku, Shibata, \&
  Taniguchi}]{sekiguchi:2016bjd}
Sekiguchi, Y., Kiuchi, K., Kyutoku, K., Shibata, M., \& Taniguchi, K. 2016,
  Phys. Rev., D93, 124046, 1603.01918

\bibitem[{Shibata {et~al.}(2017)Shibata, Fujibayashi, Hotokezaka, Kiuchi,
  Kyutoku, Sekiguchi, \& Tanaka}]{shibata:2017xdx}
Shibata, M., Fujibayashi, S., Hotokezaka, K., Kiuchi, K., Kyutoku, K.,
  Sekiguchi, Y., \& Tanaka, M. 2017, 1710.07579

\bibitem[{Siegel {et~al.}(2014)Siegel, Ciolfi, \& Rezzolla}]{siegel:2014ita}
Siegel, D.~M., Ciolfi, R., \& Rezzolla, L. 2014, Astrophys. J., 785, L6,
  1401.4544

\bibitem[{Siegel \& Metzger(2017)}]{siegel:2017nub}
Siegel, D.~M., \& Metzger, B.~D. 2017, Phys. Rev. Lett., 119, 231102,
  1705.05473

\bibitem[{Smartt {et~al.}(2017)}]{smartt:2017fuw}
Smartt, S.~J., {et~al.} 2017, Nature, 1710.05841

\bibitem[{Soares-Santos {et~al.}(2017)}]{soares-santos:2017lru}
Soares-Santos, M., {et~al.} 2017, Astrophys. J., 848, L16, 1710.05459

\bibitem[{Steiner {et~al.}(2013)Steiner, Hempel, \& Fischer}]{steiner:2012rk}
Steiner, A.~W., Hempel, M., \& Fischer, T. 2013, Astrophys. J., 774, 17,
  1207.2184

\bibitem[{Takami {et~al.}(2014)Takami, Rezzolla, \& Baiotti}]{takami:2014zpa}
Takami, K., Rezzolla, L., \& Baiotti, L. 2014, Phys. Rev. Lett., 113, 091104,
  1403.5672

\bibitem[{Tanaka {et~al.}(2017)}]{tanaka:2017qxj}
Tanaka, M., {et~al.} 2017, Publ. Astron. Soc. Jap., 1710.05850

\bibitem[{Tanvir {et~al.}(2017)}]{tanvir:2017pws}
Tanvir, N.~R., {et~al.} 2017, Astrophys. J., 848, L27, 1710.05455

\bibitem[{Troja {et~al.}(2017)}]{troja:2017nqp}
Troja, E., {et~al.} 2017, Nature, 1710.05433

\bibitem[{Typel {et~al.}(2010)Typel, Ropke, Klahn, Blaschke, \&
  Wolter}]{typel:2009sy}
Typel, S., Ropke, G., Klahn, T., Blaschke, D., \& Wolter, H.~H. 2010, Phys.
  Rev., C81, 015803, 0908.2344

\bibitem[{Villar {et~al.}(2017)}]{villar:2017wcc}
Villar, V.~A., {et~al.} 2017, Astrophys. J., 851, L21, 1710.11576

\bibitem[{Wade {et~al.}(2014)Wade, Creighton, Ochsner, Lackey, Farr,
  Littenberg, \& Raymond}]{wade:2014vqa}
Wade, L., Creighton, J. D.~E., Ochsner, E., Lackey, B.~D., Farr, B.~F.,
  Littenberg, T.~B., \& Raymond, V. 2014, Phys. Rev., D89, 103012, 1402.5156

\bibitem[{Wu {et~al.}(2016)Wu, Fernández, Martínez-Pinedo, \&
  Metzger}]{wu:2016pnw}
Wu, M.-R., Fernández, R., Martínez-Pinedo, G., \& Metzger, B.~D. 2016, Mon.
  Not. Roy. Astron. Soc., 463, 2323, 1607.05290

\bibitem[{Yang {et~al.}(2017)Yang, Paschalidis, Yagi, Lehner, Pretorius, \&
  Yunes}]{yang:2017xlf}
Yang, H., Paschalidis, V., Yagi, K., Lehner, L., Pretorius, F., \& Yunes, N.
  2017, 1707.00207

\end{thebibliography}
\end{document}